\documentclass[%
 aip,
rsi,%
 amsmath,amssymb,
 reprint,%
]{revtex4-2}

\usepackage[dvipsnames,table]{xcolor}
\usepackage{xcolor}
\usepackage{graphicx}
\usepackage{dcolumn}
\usepackage{bm}
\usepackage{nicefrac}
\newcommand{\n}[1]{\mathbf{#1}}

\begin{document}

\preprint{AIP/123-QED}

\title[Rodr\'{i}guez et al.]{Islands and current singularities in quasisymmetric toroidal plasmas.}

\author{E. Rodr\'{i}guez}
 \altaffiliation[Email: ]{eduardor@princeton.edu}
 \affiliation{ 
Department of Astrophysical Sciences, Princeton University, Princeton, NJ, 08543
}
\affiliation{%
Princeton Plasma Physics Laboratory, Princeton, NJ, 08540
}%

\author{A. Bhattacharjee}
 \altaffiliation[Email: ]{amitava@princeton.edu}
 \affiliation{ 
Department of Astrophysical Sciences, Princeton University, Princeton, NJ, 08543
}
\affiliation{%
Princeton Plasma Physics Laboratory, Princeton, NJ, 08540
}%

\date{\today}

\begin{abstract}
The presence of current singularities in a quasisymmetric magnetic field is explored. Quasisymmetry is shown effective in isolating Pfirsch-Schl\"{u}ter singularities, to leading order, to a single magnetic surface resonant with the helicity of the symmetry. The effects of pressure driven currents are analysed, indicating that exclusion of this surface from the plasma volume reduces the potential opening of islands, but does not generally eliminate them completely due to higher order asymmetric geometric effects. { These three-dimensional effects are contained in quasisymmetry and indicate the complexity of finding consistent solutions and their potential sensitivity.} The $\delta$-function current singularities show a distinct quasisymmetric behaviour only when the higher-order Fourier content of $B$ is relevant for the resistive stability parameter $D_R$ (not included in leading-order near-axis expansions). In such scenarios, quasisymmetry proves advantageous, both in simplicity and avoidance of amplification by low-order rational surfaces.
\end{abstract}

\maketitle

\section{Introduction:}\label{sec:intro}
The main purpose of magnetic confinement configurations is to confine a hot plasma within. The concept revolves around the capability of magnetic fields to control the motion of charged particles. Particles mainly gyrate around field lines while moving quickly along them. However, this leading-order behavior is not sufficient to confine a plasma for long enough to reach self-sustained thermonuclear fusion. It is generally necessary to design the magnetic configuration with some additional property that enhances its power to confine particles. \par
One such property is \textit{quasisymmetry}.\cite{rodriguez2020,boozer1983,nuhren1988,Helander2014,burby2020} Quasisymmetry (QS) may be understood to be the minimal property of the magnetic field necessary for the dynamics of charged particles to possess an additional approximately conserved momentum.\cite{rodriguez2020} This additional conservation, via a generalised version of \textit{Tamm's theorem}, prevents the particles from drifting away. The motion is instead restricted to be approximately over \textit{magnetic flux surfaces} (for thermal particles). These surfaces are defined everywhere tangent to magnetic field lines, and thus act as bounding surfaces. In toroidal geometry, it is then advantageous to construct configurations with nested magnetic surfaces. However, as is well known, it is often not possible to maintain nested magnetic surfaces.\cite{grad1967,reiman1986,hudson2010} Instead, \textit{magnetic islands}\cite{Helander2014,wessonTok} open up, especially at low-order rational surfaces. Unavoidably, these islands lead to enhanced particle and energy losses.  \par
Magnetic islands have been subject of analysis for a long time. Formally, they can be related to current singularities\cite{boozer1981,cary1985,hegna1989,reiman1984} of three-dimensional (3D) equilibrium configurations {  forced to possess nested surfaces}. This paper studies these current singularities and the potential formation of islands in {  `mostly'} quasisymmetric magnetic fields. Specifically, we address the question: does QS affect significantly current singularities and thus the potential opening of magnetic islands? \par
The paper is organised as follows. In Section II we investigate the compatibility between exact quasisymmetry and magnetic islands. The following three sections (III, IV and V) are set to analyse the appearence of current singularities in configurations that are quasisymmetric with nested flux surfaces and the opening of islands when a small but finite dissipation is introduced to amend the singularities. This is done for both flavours of existing singularities: Pfirsch-Schl\"{u}ter and $\delta$-function singularities. { Special attention is paid to the interaction between the geometry and the Pfirsch-Schl\"{u}ter currents.} We close the paper with a summary and possible extensions of our work. 


\section{Weak quasisymmetry and magnetic islands}
\subsection{Defining quasisymmetry} \label{sec:defQS}
We start this section with a rigorous definition of QS. { Quasisymmetry (in its \textit{weak} sense) is the minimal property of a magnetic field that confers the motion of the guiding centre of charged particles with an approximately conserved momentum\cite{rodriguez2020,burby2020} (to first order in gyro-radius).} This physical definition may be formalised in a number of ways.\cite{rodriguez2020,rodriguez2020i} One such form, perhaps the most natural one\cite{rodriguez2020,burby2019}, ties the existence of a non-trivial vector field $\n{u}$ that satisfies
\begin{gather}
    \n{u}\cdot\nabla B=0, \label{eqn:QSa}\\
    \n{B}\times\n{u}=\nabla\Phi, \label{eqn:QSb}\\
    \nabla\cdot\n{u}=0, \label{eqn:QSc}
\end{gather}
to a \textit{quasisymmetric} magnetic field $\n{B}$. It is clear from Eq.~(\ref{eqn:QSa}) that the vector field $\n{u}$ represents the direction of symmetry of $|\n{B}|$. This vector lives on magnetic flux surfaces labelled by a single-valued function $\Phi$ according to Eq.~(\ref{eqn:QSb}). { This would generally lead to the conclusion that $B=B (\Phi)$, the isodynamic condition, which is much too constraining for toroidal confinement.\cite{palumbo1968} If we are to avoid this conclusion---that the magnetic field magnitude must be constant on flux surfaces---we must require the vector field $\n{u}$ have rational helicity and thus close on itself everywhere within the configuration.} Moreover, as the magnitude of $\n{B}$ dictates the direction of $\n{u}$ and it is a physical quantity, it should only vary smoothly {  from surface to surface. The only way to satisfy both conditions is for the symmetry vector $\n{u}$ to have a single rational helicity globally.}
\par
In this picture of QS, { we have invoked no consideration of macroscopic equilibrium.} It is customary (and physically motivated), however, to assume magnetohydrostatic (MHS) equilibrium in the simplest form $\n{j}\times\n{B}=\nabla p$, where $\n{j}$ is the plasma current and $p$ is the isotropic plasma pressure. In Sections III, IV and V of the present paper we shall also adopt this simple form of equilibrium in order to make clear comparisons with the existing literature. However, it is worth noting that recent research points towards force balance with more general forces (in particular anisotropic pressure) as a more natural framework in which to fit quasisymmetric fields\cite{rodriguez2020d,rodriguez2020i,rodriguez2020ii,constantin2020}. An analysis of singularities in the context of more general forces is left for future work. \par
In MHS, the existence of Boozer coordinates\cite{boozer1981} $\{\psi,\theta,\phi\}$ (possible as a result of $\n{j}\cdot\nabla\psi=0$) allows for the use of the so-called \textit{Boozer formulation} of QS.\cite{nuhren1988,boozer1995,Helander2014} In this form, a field is quasisymmetric if the Fourier content of its local magnetic energy $B^2$ has a single helicity $\chi=\theta-\tilde{\alpha}\phi$ with $\tilde{\alpha}\in\mathbb{Q}$. This is equivalent to saying that the Jacobian $\mathcal{J}=(G+\iota I)/B^2$ has a single helicity, where $G$ and $I$ are the Boozer currents and functions of $\psi$. {  In the context of MHS the \textit{weak} form of QS in Eqs.(\ref{eqn:QSa})-(\ref{eqn:QSc}) is equivalent to this Boozer formulation\cite{rodriguez2020i,rodriguez2020e}, generally referred to as (plain) QS.} \par  

\subsection{Describing magnetic islands}
\begin{figure}
    \centering
    \includegraphics[width=0.5\textwidth]{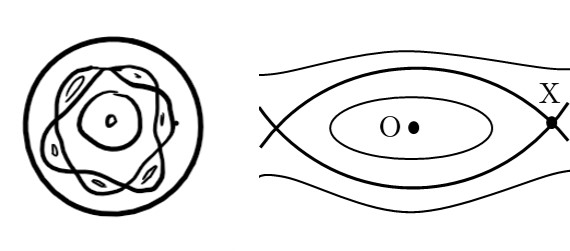}
    \caption{\textbf{Geometry of magnetic island.} (Left) A depiction of a magnetic island chain in a Poincar\'{e} section of a toroidal device. (Right) A sketch of a magnetic island where the horizontal direction represents $\alpha=\theta-\iota\phi$ (the helical magnetic angle) and the vertical direction $x$. The O and X points are shown, the latter corresponding to $\alpha=\pi/2m$.}
    \label{fig::magIslGeo}
\end{figure}

We now introduce the structure of magnetic islands. We shall do this in its most basic form, and refer the reader to the literature for more extensive discussions.\cite{Helander2014,wessonTok} We start with a magnetic field, $\n{B}_0$, with well-defined nested flux surfaces. We then choose a rational surface $\psi=\psi_r$ and allow for an additional, small \textit{resonant} magnetic field $\n{B}_1$. By \textit{resonant} we mean that $\n{B}_1$ shares the angular dependence of the rational surface with $\iota_r=n/m$ (where $n,m\in\mathbb{Z}$) and has a non-zero component normal to the surface. As a result of this \textit{resonant} field, the flux surface structure associated to $\n{B}_0+\n{B}_1$ near the original resonant surface is modified, breaking the global nested structure. Following the non-canonical perturbation theory of Cary and Littlejohn\cite{cary1983,cary1985}, {  these new surfaces are described by the modified surface label $\bar{\psi}^*$, defined as $\Bar{\psi}^*=\frac{1}{2}\iota_r'x^2-A_0\cos m(\theta-\iota\phi)$, where $x=\psi-\psi_r$, with the perturbation of the field chosen to be the toroidal vector potential $A_0\cos m(\theta-\iota\phi)$.} Typical representation of a magnetic island chain is shown in Fig.~\ref{fig::magIslGeo}.  \par
The island structure has two differential features. The O point is a closed rational magnetic field line, similar to the magnetic axis. The separatrix is the other important feature of islands, separating them from the remainder of the configuration. An important feature of the island is its width in the radial direction. This width can be related directly to the radial component of the resonant magnetic field $B_{1\rho nm}$. The island width $W_i$ is,\cite{wessonTok,reiman1984,bhattacharjee1995}
\begin{equation}
    W_i\approx4\sqrt{\frac{\rho\iota}{m\iota'}\frac{|B_{1\rho nm}|}{B_\theta}}, \label{eqn:islWidthB}
\end{equation}
where $m$ is the poloidal mode number, $\rho$ the minor radius and $B_\theta$ the poloidal component of the magnetic field. Here we are assuming surfaces to be approximatedly cylindrical.

\subsection{Compatibility between QS and islands}
{  With the main concepts established, let us ask two questions. First: can we have magnetic island chains in a configuration that is truly quasisymmetric everywhere? We will see that quasisymmetric configurations are largely incompatible with the presence of magnetic islands (this section). { Second: If a "mostly quasisymmetric" field is considered, what controls the size of islands?} The idea for the latter is to, first, evaluate the presence of current singularities and return currents when imposing QS and MHS. Then, to consider the associated resonant magnetic perturbation that these currents are responsible for, and estimate the finite island that would open if some small but finite resistivity is allowed to regularize the singularities and provide access to magnetic islands. This last scenario we refer to as mostly quasisymmetric, { as it originates in a field that is a quasisymmetric magnetic field plus a perturbation that breaks QS. More details will be provided in Sections III-V.}  \par
As mentioned above, we deal with the first question in this section. Let} us assume we have an island chain at some resonant surface with $\iota=\iota_1$. If we enforce QS exactly and demand $B\neq0$, then at the O-point the second of the quasisymmetric conditions Eq.~(\ref{eqn:QSb}) must hold. This implies that the magnetic field magnitude $|\n{B}|$ is constant along the {  field} line joining the O-points. Because $\n{u}$ traces streamlines of constant $B$, it must trace the O-point. Because $\n{u}$ must have a single global rational helicity, then this value must match $\iota_1$.  \par
Of course, if the above is true for one island chain, it must break for any other island with a different helicity. Thus a magnetic configuration can only coexist with an island resonant with the direction of symmetry. To allow for multiple island chains, either $|\mathbf{B}|$ must be a flux function at a flux surface somewhere in the configuration or QS is not exactly satisfied. From this discussion it is clear that the QS property confers a resilience to magnetic islands, and it does so independent of the form of force balance. {  If we strictly impose QS no islands can appear in the problem (except possibly at $\iota=\tilde{\alpha}$). } \par

\section{Current singularities: a primer}
Let us consider the `mostly' quasisymmetric scenario introduced above. That is to say, consider some quasisymmetric magnetic field $\n{B}_0$ that has well-defined, nested flux surfaces by construction. In addition, let it be in MHS force-balance. Generally, forcing the system to have nested toroidal flux surfaces in equilibrium leads to the formation of current singularities\cite{boozer1981,cary1985,hegna1989,reiman1984}. 
We explore these now. \par
The appearence of current singularities is easily recognised from studying the current continuity equation\cite{boozer1981,reiman1984,bhattacharjee1995}. To prevent charge from piling up, the current must satisfy $\nabla\cdot \n{j}=0$. From MHS force balance, the component of the current density perpendicular to the magnetic field is known to be $\n{j}_\perp=\n{B}\times\nabla p/B^2$. To find the parallel piece of the current (which does not contribute to force balance) a magnetic differential equation $\n{B}\cdot\nabla\sigma=-\nabla\cdot\n{j}_\perp$ needs to be solved. Here $\sigma=\n{j}\cdot\n{B}/B$. The solution for $\sigma$ of this equation\cite{boozer1981,bhattacharjee1995} is straightforward when {  Fourier expanded} using Boozer coordinates\cite{boozer1983} (not including $n=m=0$),
\begin{equation}
    \sigma_{nm}=\frac{p'}{n/m-\iota}\frac{G+(n/m)I}{G+\iota I}\mathcal{J}_{nm}+\hat{\sigma}_{nm}\delta(\psi-\psi_{nm}). \label{eqn:currSingEq}
\end{equation}
The subscripts $n,~m$ label the Fourier mode of the corresponding function except for $\psi_{nm}$, for which it is a label of the magnetic flux surface resonant with the $(n,m)$ mode. \par
Equation (\ref{eqn:currSingEq}) has two separate contributions. The first one is the explicitly pressure dependent piece{, often called \textit{Pfirsch-Schl\"{u}ter} current. This term driven by the force balance } shows a denominator that vanishes when the mode $(n,m)$ matches the rotational transform of the field. This resonance is commonly referred to as the \textit{Pfirsch-Schl\"{u}ter} (PS) or $1/x$ singularity. The second term is referred to as a $\delta$-\textit{function current singularity}. This is the solution to the homogeneous part of the magnetic differential equation, and spikes at rational values of the rotational transform. Unlike the PS singularity which is not integrable and can exhibit logarithmic divergence, the $\delta$-function singularity is integrable. {The net parallel current in the problem is controlled by the $n=0=m$ piece of $\sigma$. This includes a smooth flux function term as part of the homogeneous solution which will be determined later. }\par
Of course, the presence of singularities in the problem is not physical. It is a signal that the ideal MHD model with the assumption of nested surfaces is being pushed into a space where the validity of the model is questionable. {  To evaluate how this model breaks down, it is common to treat the problem perturbatively. We assume the ideal MHD solution with nested surfaces as a leading order $\mathbf{B}_0$, and compute the currents from the model. The non-physical singularity from the PS current is customarily avoided demanding $p'=0$ at resonance. The remaining non-singular PS current can lead to an effective resonant magnetic perturbation $\n{B}_1$, one which can potentially lead to to the opening of islands. (These islands did not show up in $\mathbf{B}_0$ because of the nested surface assumption.) This scenario is explored in Sec.~IV for a field $\n{B}_0$ with QS properties. The $\delta$-function current singularities appear in the problem to guarantee that $\mathbf{B}_0$ has nested surfaces as it has been assumed. To find the amplitude and resolve these singularities, we need to include additional physics and allow for the opening of magnetic islands. This part of the problem is treated in Section V, where the $\delta$-function part is (somewhat artificially) separated from the PS part. } The presented considerations do not regard finite vacuum island widths nor the full interaction between the two types of resonances, for which an extension would be required.\cite{bhattacharjee1995} 

\section{Pfirsch-Schl\"{u}ter current singularity and quasisymmetry} \label{sec:PS}
To understand the influence of QS on the Pfirsch-Schl\"{u}ter currents { (and vice-versa)}, and the potential appearence of islands, we separate the problem into two steps. The first one is the evaluation of the currents themselves. But knowledge of the current does not suffice to estimate the potential opening of magnetic islands. To complete the gap a second step is needed in which the magnetic field due to these currents is computed. We shall address both of these aspects in this section.

\subsection{PS current singularity with QS} 
The first step in the problem is to understand how the current singularity is affected by QS. \par

Some of the factors in the singularity expression in Eq.~(\ref{eqn:currSingEq}) (e.g. $p'$ or $G$) are general and do not acquire a special form for QS. On the other hand, the behaviour of the Jacobian factor $\mathcal{J}_{nm}$ is special for a quasisymmetric configuration. To see this, recall from Sec.~\ref{sec:defQS} that a field that is quasisymmetric has a Jacobian with a single helicity. If we take $\tilde{\alpha}$ as that helicity, it follows that $\mathcal{J}_{nm}\neq0$ if $n/m=\tilde{\alpha}$. A general stellarator configuration will generally possess many non-zero $\mathcal{J}_{nm}$. \par
Therefore,
\begin{equation}
    \sigma^\mathrm{PS}=\frac{p'}{\tilde{\alpha}-\iota}\tilde{\mathcal{J}}, \label{eqn:PScurrSing}
\end{equation}
where $\Tilde{\mathcal{J}}=\mathcal{J}-\mathcal{J}_{0}$. We have assumed here the toroidal Boozer current and $\langle\mathbf{j}\cdot\mathbf{B}\rangle=0$ vanish, which is a typical assumption for stellarators. This choice reflects the freedom we had when inverting the magnetic differential equation for $\sigma$. \par 
The current singularity is localized and occurs at a single rational surface. That surface matches the helicity of the assumed symmetry. This allows for a simple exclusion of the singularity from the plasma volume, by construction. {  In the case of quasiaxisymmetric configurations (an example of which is the tokamak), the singularity is localised at $\iota=0$. Thus, it is easily kept outside the plasma volume so long as one has a finite rotational transform. For the same reason, quasi-helically symmetric devices require a more careful design to avoid this singularity. This is especially true for the lowest helicities, $\tilde{\alpha}$, which could match the value of the rotational transform within the plasma body.} The need to consider a single surface makes the design process straightforward (at least in principle) compared to the general 3D situation, in which one needs to worry about a larger number of resonant rational surfaces. \par
The changes in the behaviour of PS current singularities are governed by the spectral simplicity of $B$. In that sense, there is no qualitative difference  between a fully symmetric configuration (like axisymmetry) or a truly 3D-geometry quasisymmetric one. {  To pick on additional differences, one is obliged to consider additional aspects of the geometry. Using the Jacobian $\mathcal{J}$ already introduced some geometrical aspects into the problem. But this scalar quantity does not include a description of the full geometry. Most notably{, for QS,} it does not include information about the \textit{asymmetric} aspects of the geometry, {which do play a prominent role on the resonant magnetic fields associated with the PS currents.} We now study these potentially resonant, island-opening perturbations due to the PS current taking the geometry into account}. { We do so by considering particularly simple forms of geometry as an external element to the problem. This should provide insight on the role of PS currents and the importance of geometry. The full, self-consistent problem would not treat the geometry as an external element, but would rather consider the geometry of $\mathbf{B}_0$. We do not attempt this here.}

\subsection{Direct resonance: 'circular' geometry, a recap}
To see how to incorporate the geometric aspects into the calculation, we start by considering the simplest case of circular cross-section. We set up the problem by asking what the resonant magnetic fields $\n{B}_{nm}$ are due to the PS currents corresponding to the QS field $\n{B}_0$ in circular geometry. Generally, the resonant PS currents, and through the resonant $\mathbf{B}_{nm}$ piece, give rise to finite islands incompatible with $\mathbf{B}_0$ (if not prevented by the presence of $\delta$-function current singularities for now ignored), whose width can be estimated applying Eq.~(\ref{eqn:islWidthB}). The perturbative procedure presented here follows that of [\onlinecite{reiman1984}], and thus we refer to it for more details on the derivation. Here we sketch the main steps, important to set the stage for the implications of a more complicated shaping. \par
Consider a magnetic field $\n{B}_0$ that is quasisymmetric (so that $B_0=B_0(\psi,\theta-\tilde{\alpha}\phi)$ can be taken), in MHS equilibrium with nested surfaces. Sticking for simplicity to the assumption $I=0$ (no net toroidal current) for $\n{B}_0$, the field may be written in Boozer coordinates (defined with respect to $\n{B}_0$),
\begin{equation}
    \n{B}_0=G(\psi)\nabla\phi+B_\psi\nabla\psi=\nabla\psi\times\nabla\theta+\iota\nabla\phi\times \nabla\psi. \label{eqn:fieldCoCon}
\end{equation}
We shall assume that the resonant piece $\n{B}_{nm}$ due to the resonant current is small and localised to the resonant flux surfaces. We ignore its possible contribution to force-balance (possible variations of $p$ over surfaces due to islands as explored in [\onlinecite{reiman2016}] are not considered here neither). It is straightforward in this case to write down the current associated to the problem using Eq.~(\ref{eqn:PScurrSing}),
\begin{align}
    \n{j}=\frac{p'}{\Tilde{\alpha}-\iota}\Tilde{\mathcal{J}}\n{B}_0+\frac{p'}{B_0^2}\n{B}_0\times\nabla\psi.
\end{align}
($\delta$-function singularities have been dropped here.) Using the Boozer and contravariant forms of the field in Eq.~(\ref{eqn:fieldCoCon}), the expression for the current density can be written explicitly as the curl of a vector field $\n{H}$,\cite{reiman1984}
\begin{multline}
    \n{H}=-\left(\int\mathcal{J}_{0} p'\mathrm{d}\psi\right)\nabla\phi+\\
    +{\sum_n}'\frac{p'}{\Tilde{\alpha}-\iota}\frac{1}{inM}\mathcal{J}_{n}e^{inM(\tilde{\alpha}\phi-\theta)}\nabla\psi, \label{eqn:Hfield}
\end{multline}
where for brievity of notation we have shortened $\mathcal{J}_{nN,nM}=\mathcal{J}_n$ where $\tilde{\alpha}=N/M$ and excluded $\mathcal{J}_{0}$ from the sum. Of course, inversion of the curl operation that led to Eq.~(\ref{eqn:Hfield}) is not unique. This leaves some gauge freedom in the form of a scalar field $F$ such that $\n{B}=\n{H}+\nabla F$. \par
To find an expression for $F$ we exploit $\nabla\cdot\n{B}=0$,
\begin{equation}
    \nabla^2F=-\nabla\cdot\n{H}, \label{eq:poissRes}
\end{equation}
which results in a Poisson-like equation. This equation is to be solved for $F$ subject to the condition that $F$ vanishes away from the plasma. \par
To proceed further analytically with this equation, we shall consider a convenient set of quasi-cylindrical coordinates: the orthogonal Mercier set.\cite{Helander2014} The set consists of coordinates $\{\rho,\omega,l\}$, where $\rho$ is a radial cylindrical coordinate in the plane perpendicular to the axis at each $l$, $l$ parametrises the distance along the magnetic axis and $\omega$ is a helical redefinition of the cylindrical angle. \par
Assuming a large aspect ratio $\kappa\rho\ll1$ ordering {  (where $\kappa$ is the curvature of the magnetic axis)}, the Laplacian and divergence operators can be written in a straightforward way. Dropping the toroidal derivatives under the assumption of the large aspect ratio, the relevant differential operators in Eq.~(\ref{eq:poissRes}) read, 
\begin{equation}
    \nabla\cdot\n{H}\approx\frac{1}{\rho}\partial_\rho(\rho H_\rho)+\frac{1}{\rho}\partial_\omega H_\omega, \label{eqn:divH}
\end{equation}
and
\begin{equation}
    \nabla^2 F\approx\frac{1}{\rho}\partial_\rho\left(\rho\partial_\rho F\right)+\frac{1}{\rho^2}\partial_\omega^2F. \label{eq:laplcianMercer}
\end{equation}
Even though these operators acquire a convenient form in Mercier coordinates, the problem is not yet solved. To compute the resonant piece of $\n{B}$ needed for Eq.~(\ref{eqn:islWidthB}) we need to Fourier decompose $F$ and the components of $\n{H}$, and do so in flux coordinates. Therefore, it is necessary to know the mapping between Mercier and Boozer coordinates. It is here that the geometry of the problem makes a difference. { Here we shall artificially decouple the geometry of flux surfaces from the field $\mathbf{B_0}$, and consider circular cross-sections.\cite{reiman1984}} \par
With this perspective, 
\begin{gather}
    l\approx L\phi/2\pi, \label{eqn:MercBoozPhiC}\\
    \omega\approx\alpha, \label{eqn:MercBoozOmegC}\\
    \psi\approx\frac{B_{l0}}{2}\rho^2, \label{eqn:MercBoozPsiC}
\end{gather}
near the magnetic axis, where $\alpha=\theta-\iota\phi$. {  The meaning of $B_{l0}$ and the details of this construction are detailed in Appendix A.} \par
With these relations we may try to complete all the pieces needed to solve Eq.~(\ref{eq:poissRes}) for $F$. From Eq.~(\ref{eqn:Hfield}) it follows that $H_\omega=0$ and
\begin{equation}
    H^\circ_{\rho nm}=B_{l0}\rho\frac{p'}{\Tilde{\alpha}-\iota}\frac{1}{inM}\mathcal{J}_{n}\delta_{n,m\tilde{\alpha}}, \label{eqn:Hcirc}
\end{equation}
where ${}^\circ$ represents the circular surface case and $\delta$ in the expression represents a Kronecker delta. \par
Fourier-decomposing Eq.~(\ref{eq:poissRes}) we obtain 
\begin{equation}
    \frac{1}{\rho}\partial_\rho(\rho\partial_\rho F_{nm})-\frac{m^2}{\rho^2}F_{nm}\approx-\frac{1}{\rho}\partial_\rho(\rho H^\circ_{\rho nm}), \label{eqn:possFourCirc}
\end{equation}
which can be solved straightforwardly by variation of parameters\cite{reiman1984}. It is most important to note that the source term of Eq.~(\ref{eqn:possFourCirc}) vanishes for modes that are not resonant with the QS helicity. Thus, when computing the resonant piece $B_{\rho nm}=\partial_\rho F_{nm}+H_{\rho nm}$ that could potentially lead to the opening of a magnetic island, it is clear from Eqs.~(\ref{eqn:possFourCirc}) and (\ref{eqn:Hfield}) that a resonant magnetic island may only possibly open at the QS-resonant surface. So as discussed earlier, excluding this surface would prevent the island due to the resonant PS current. \par
The fact that the leading order appearence of islands is so localised offers a control that could be practically exploited. Perhaps this control could be used in quasi-helically symmetric (QHS) configurations to naturally have an island divertor at finite plasma $\beta$. One could devise it by bringing the resonant surface close to the edge $\iota$. This idea is not as natural in quasiaxisymmetric (QAS) configurations, given that $\iota$ generally monotonically increases with radius and the resonant helicity is $\tilde{\alpha}=0$. The island divertor design for a QAS configuration will require a more careful design (possibly more QS degrading {  as one will have an island chain appearing at a rational surface other than the symmetry resonant one}). However, such designs do exist. \par
{{The `mostly' QS property of $\mathbf{B}_0$ has almost exclusively been encoded in this approach in the symmetric content of the Jacobian. And we saw the important localisation consequences of it.} The details of the Fourier content of that symmetric part have however been dealt with in most general terms. { Given the importance of $\mathcal{J}$, a particular design of the mode content could however modify the observations above.} In fact, to eliminate the presence of the potential resonant island, one would have to construct the symmetry-resonant surface in a way that the field would be isodynamic. I.e., the magnetic field strength would have to be constant over the surface, { which would avoid the bad behaviour of $\mathbf{B}_0$ around the symmetry resonant surface.} \par
In practice, the system will not be exactly quasisymmetric. Deviations from QS will bring general 3D configuration features back into the problem. In particular, current singularities will appear in additional resonant surfaces. This could lead to additional islands of significant width, especially at low order rationals. The size of these islands could then be a physical measure of departure from QS. This is an example illustrating how QS may be `measured' in different ways, depending on the focus of the physics of interest. This should be taken into consideration when optimising stellarators for QS.

\subsection{Resonant sidebands: elliptical surfaces}
The simplicity of Mercier coordinates in circular geometry has led us to a clear understanding of resonant mode behavior. Only the mode resonant with the QS helicity is excited. We only need to worry about islands that arise precisely at the QS resonant surface. This is in fact the same situation that occurs in axisymmetric devices, in which the current singularities are potentially deleterious at the directly resonant surfaces. \par
The story changes when the geometry of the problem is more complicated. And the geometry is generally more complicated in stellarators. { To understand how geometry affects the picture, we adopt} the most natural generalisation of a circular cross-section: a mild { rotating} ellipse. This is a more realistic situation\cite{garrenboozer1991a,landreman2018a,rodriguez2020ii} {  and provides an explicitly asymmetric piece to the problem. { Of course, this still constitutes a significant simplification away from the full $\mathbf{B}_0$ shaping, especially as we consider shaping parameters as external controls.}} Let us describe the elliptical shape of the surfaces as the loci $(x,y)=(re^{-\eta/2}\cos v,re^{\eta/2}\sin v)$ in the plane perpendicular to the magnetic axis, where $\eta$ is a measure of the ellipticity, $r$ is a `radial' coordinate and $v$ is an angular parametrisation of the ellipse. By mild ellipticity we mean that $\eta\ll 1$.  \par
With this new shaping, the mapping of coordinates close to the magnetic axis changes quite significantly. The details of the construction are included in Appendix A. Here we simply collect the leading order coordinate mapping,
\begin{gather}
    l\approx L\phi/2\pi, \label{eqn:MercBoozPhi}\\
    \tan\left[\omega+d-\int_0^l\tau\mathrm{d}l'\right]\approx e^\eta\tan\left[\alpha+\int_0^l\frac{d'-\tau}{\cosh\eta}\mathrm{d}l'\right], \label{eqn:MercBoozOmeg}\\
    \psi\approx \frac{B_{l0}}{2}\rho^2(e^\eta\cos^2 u+e^{-\eta}\sin^2 u), \label{eqn:MercBoozPsi}
\end{gather}
where $u=\omega+d(l)-\int_0^l\tau\mathrm{d}l'$, $d(l)$ represents the rotation angle of the elliptical cross section measured between the curvature vector and the ellipse's minor axis (considered as an externally controlled quantity for the analysis), $\tau$ is the torsion of the magnetic axis, and $B_{l0}$ is the magnetic field magnitude on axis. It will prove convenient to define $\bar{\alpha}=\alpha+\int_0^l(d'-\tau)/\cosh\eta\mathrm{d}l'$.  { The shaping parameters $\eta$ and $d$ describe the asymmetric geometry and are considered as external parameters.}\par
It is clear from the coordinate mapping defined in Eqs.~(\ref{eqn:MercBoozPhi})-(\ref{eqn:MercBoozPsi}) that as a result of the elliptical shaping the mapping loses the orthogonality of Eqs.~(\ref{eqn:MercBoozPhiC})-(\ref{eqn:MercBoozPsiC}). The `poloidal' angular coordinate $\omega$ now generally reads $\omega=\omega(\alpha,\phi)$ from Eq.~(\ref{eqn:MercBoozOmeg}). The flux $\psi$, in turn, is no longer just a relabelling of the radial coordinate, as it includes some angular variation in Eq.~(\ref{eqn:MercBoozPsi}). Although we have this additional mixing of coordinates, it is still true that $\partial_\rho h(\theta,\phi)=0$ for any function $h$. \par
Due to this additional complication, the trivial mode resolution of Eq.~(\ref{eqn:possFourCirc}) in the circular case is no longer so. {  A detailed construction is presented in Appendix B. Here we present the key pointers.} The idea is to get to the equivalent of Eq.~(\ref{eqn:possFourCirc}) for a mildly elliptical geometry. \par
We start from the relevant Mercier projections of $\n{H}$ needed to compute $\nabla\cdot\mathbf{H}$. We may write them explicitly using Eq.~(\ref{eqn:MercBoozPsi}),
\begin{gather}
    H_\rho=H_\rho^\circ\left(\cosh\eta+\sinh\eta\cos2u\right), \label{eqn:modHrho} \\
    H_\omega=-H_\rho^\circ\sinh\eta\sin2u.
\end{gather}
Then, including and keeping terms only to leading order in $\eta$, we may compute the two pieces in Eq.~(\ref{eqn:divH}), with special care taken for the new terms arising from $\partial_\omega H_\omega$,
\begin{equation}
    \nabla\cdot\n{H}\approx\frac{1}{\rho}\partial_\rho\left(\rho H_\rho^\circ\right)-\mathcal{H} e^{2i\bar{\alpha}}, \label{eqn:divHEllip}
\end{equation}
where $\mathcal{H}\approx iB_{l0}p'\mathcal{J}\sinh\eta/(\tilde{\alpha}-\iota)$.
To obtain this approximate form we assumed $p'/(\tilde{\alpha}-\iota)$ to be roughly constant and $\mathcal{J}_{nm}$ to scale like $\propto\psi^{m/2}$.
Comparing Eq.~(\ref{eqn:divHEllip}) to the circular problem in Eq.~(\ref{eqn:divH}), there exists an additional term proportional to $\mathcal{H}$ with an additional angular component $\bar{\alpha}$. This is indicative of mode coupling. {  Relaxation of certain assumptions detailed in the Appendix will change the exact form of $\mathcal{H}$ and lead to the appearence of additional coupling terms. However, the qualitative picture does not change.} \par
A similar analysis can be carried out for the Laplacian piece too (see Appendix B). Resolving $F$ in a Fourier series in $(\theta,\phi)$ like for the circular case, defining the pseudo-radial variable $\bar{\rho}=\sqrt{2\psi/B_{l0}}$ and looking at each of the $(n,m)$ terms in the Fourier series definition of $F$,
\begin{equation}
    \nabla^2 F\approx\frac{\bar{\rho}^2}{\rho^2}\frac{1}{\bar{\rho}}\partial_{\bar{\rho}}(\bar{\rho}\partial_{\bar{\rho}} F_{nm})+\frac{1}{\rho^2}\hat{\Xi} F _{nm}, \label{eqn:lapFell}
\end{equation}
where $\hat{\Xi}=-m^2+e^{2i\bar{\alpha}}\hat{\Xi}_2+e^{-2i\bar{\alpha}}\hat{\Xi}_{-2}$ and the hats denote that they are operators. In particular,
\begin{gather}
    \hat{\Xi}_2\approx m(m-1)\sinh\eta+2\psi(m-1)\sinh\eta\partial_\psi, \\
    \hat{\Xi}_{-2}\approx m(m+1)\sinh\eta-2\psi(m+1)\sinh\eta\partial_\psi.
\end{gather}
Note that once again, as for $\nabla\cdot\mathbf{H}$, we obtain mode coupling. The flux function derivatives do not introduce additional angular dependence (they act on $F_{nm}$ which is by construction a flux function). The sideband contributions have a particular simple form for those modes with $m=0,\pm1$. \par
With the Laplacian and divergence pieces in place, a mode-resolved Poisson equation like Eq.~(\ref{eqn:possFourCirc}) can now be obtained. There are two important steps. First, and because the Fourier components are defined in angular Boozer coordinates, we need to find the correct form for $\bar{\alpha}$ in Boozer form. Using the expression for the rotational transform on axis in terms of the geometry (see [\onlinecite{Helander2014}], Eq.~(44)), we may write $\bar{\alpha}=\theta-n_d \phi/2+f(\phi)$. Here $n_d\pi=\int_0^ld'\mathrm{d}l'+2\pi\tilde{\alpha}$ (effective rotation of the elliptical cross section) and $f(\phi)$ is a periodic function that may be found as an integral along the magnetic axis from the full definition of $\bar{\alpha}$. For simplicity, we shall assume $f=0$. The second important step comes from realising that Eq.~(\ref{eqn:lapFell}) is mostly a function of $\psi$ with the addition of the sideband couplings but for the factor $\bar{\rho}^2/\rho^2$. To resolve this mode equation properly we need to multiply through by its inverse (and so do the same with $\nabla\cdot\mathbf{H}$). This gives rise to additional couplings, which lead to a mode-resolved Poisson equation,
\begin{multline}
    \frac{1}{\bar{\rho}}\partial_{\bar{\rho}}(\bar{\rho}\partial_{\bar{\rho}} F_{nm})+\frac{1}{\bar{\rho}^2}\left[-m^2+e^{2i\bar{\alpha}}\hat{\Xi}_2+e^{-2i\bar{\alpha}}\hat{\Xi}_{-2}\right]F _{nm}\approx\\
    \approx\frac{1}{\bar{\rho}}\partial_{\bar{\rho}}(\bar{\rho}H_{\bar{\rho}}^\circ)+\mathcal{H}_2e^{2i\bar{\alpha}}+\mathcal{H}_{-2}e^{-2i\bar{\alpha}}, \label{eqn:coupModePoiss}
\end{multline}
where $\mathcal{H}_2$ and $\mathcal{H}_{-2}$ are once again geometry dependent pieces proportional to $\sinh\eta$ (see Appendix B for a fully fleshed form), and $H_{\bar{\rho}}^\circ=(\bar{\rho}/\rho)H_{{\rho}}^\circ$. \par
Equation (\ref{eqn:coupModePoiss}) constitutes a system of coupled ordinary differential equations for the unknowns $F_{nm}$. All the terms in it are flux functions, except the expicitly written coupling exponential factors. These couple the equations for the different modes; the $\hat{\Xi}$ can be seen as coupling side-band terms. To study the implications of this system of equations, we carry out a perturbative analysis of Eq.~(\ref{eqn:coupModePoiss}) exploiting the ordering assumed for the shaping parameter $\eta$. \par
We start by taking the leading order, $\eta=0$ form of Eq.~(\ref{eqn:coupModePoiss}). As required, the equation reduces to Eq.~(\ref{eqn:possFourCirc}), and thus only when the $n,m$ mode is resonant with the helicity of the QS do we get a contribution to the resonant field. The finite $\eta$ corrections will modify this expression slightly, but this is not our main concern here. It is most interesting to investigate the behaviour of those modes that in the circular limit were not excited, but may be so here. These different modes are resonant at different surfaces and thus could lead to additional magnetic island openings. \par
As an example of one such mode, focus on the $(n+n_d,m+2)$ mode, where $(n,m)$ represents a mode resonant with the QS direction. { We are for simplicity assuming that $\eta$ is roughly constant.} Its leading order equation reads,
\begin{multline}
    \frac{1}{\bar{\rho}}\partial_{\bar{\rho}}(\bar{\rho}\partial_{\bar{\rho}} F_{n+n_d,m+2})-\frac{(m+2)^2}{\bar{\rho}^2}F_{n+n_d,m+2}+\\
    +\frac{1}{\bar{\rho}^2}\hat{\Xi}_{-2}F_{nm}\approx\mathcal{H}_{-2}. \label{eqn:poissEllip}
\end{multline} 
In Eq.~(\ref{eqn:poissEllip}) the piece $F_{nm}$ resonant with QS and $\mathcal{H}_{-2}$ serve as a source for $F_{n+n_d,m+2}$, which to 0th order in the shaping vanishes. The driving of this new mode (call it a sideband) will be proportional to $\hat{\Xi}_{-2},\mathcal{H}_{-2}\propto\sinh\eta$, and thus contributes to a potential finite island opening at the rational surface $\iota=(n+n_d)/(m+2)$. \par
In Appendix C we present a formal solution for the sideband field in Eq.~(\ref{eqn:poissEllip}) and the construction of the leading order resonant field $B_{1\bar{\rho}n+n_d,m+2}$. From this resonant field one may estimate the width of potential islands using Eq.~(\ref{eqn:islWidthB}) to leading order in $\eta$. Doing so shows that the island width scales at the $\iota=(n+n_d)/(m+2)$ sideband as $W_i\sim\sqrt{p'\sinh\eta\mathcal{J}_m/\iota'|\tilde{\alpha}-\iota|}$. A similar situation occurs for the $n-n_d,m-2$ mode. {  The precise expression for the size of the island will be different from the expression for the $n+n_d,m+2$ mode, but will share the main qualitative dependence. We do not include it here.} \par
The potential island width due to the PS current is then proportional to the strength of the shaping. It is amplified as the symmetry-resonant surface is approached and is proportional to both the non-resonant Jacobian as well as the pressure. Perhaps the most important of its properties is that these potential islands will open at rational surfaces other than the QS-resonant one. { Thus, it is not enough to avoid the direct-resonant surface to prevent potential opening of islands.} The PS currents will generally lead to finite resonant fields at many rational surfaces. These surfaces belong to a selected subset. Let us delve into the structure of these sideband modes.
\par
Consider a quasisymmetric configuration with helicity $\tilde{\alpha}$. From the coupling obtained above, first-order finite width islands may appear at surfaces with $\iota_r=(m\tilde{\alpha}\pm n_d)/(m\pm 2)$ for any $m$. This constitutes, for $m>0$, a whole set of potentially problematic surfaces. For large $m$, even though it might be the case that there is a first-order (in $\eta$) non-resonant coupling, the $m$-th Fourier component of the Jacobian decays exponentially. Thus, the { lower-rational} surfaces are most important. These sidebands at first-order could couple together to give second-order sidebands and beyond. This additional coupling could be explored including higher order effects in $\eta$ in Eq.~(\ref{eqn:poissEllip}). These secondary islands would involve additional powers of $\sinh\eta$.  If the shaping became strong, the structure of the mode coupling would become more complicated. It would do so not necessarily in its mode content, but in modifying the hierarchy of island widths. This is so because the set of Poisson mode equations would probably not be solvable by standard perturbation theory.This situation is not addressed here. Relaxing some of the approximations in the text such as dropping $\partial_l$ terms will also introduce additional effects. An example of collections of potentially relevant sidebands is shown in Fig.~\ref{fig:coupEllipRat}. \par

The inclusion of additional geometric effects will lead to the appearence of even more sidebands. We do not show this explicitly, but one expects to find that the way that elliptical shaping leads to $m\pm2$ mode coupling, higher order shaping such as triangularity will couple additional modes $m\pm k$ (see Fig.~\ref{fig:coupEllipRat}). The toroidal mode coupling $n\pm n_d$ is also generally not as simple as shown here. Toroidal variations in $f(l)$ for $\bar{\alpha}$ will, by a Jacobi-Angers expansion, couple more toroidal modes. This increases the number of susceptible rational locations. { The parameter $\eta$ does also generally vary with $l$, and thus would also modify the toroidal coupling.} The more symmetric the geometry, the less important this resonance bleeding is. Therefore, there is a big difference to be expected between a general 3D quasisymmetric and {  an axisymmetric} configuration (a quantitative comparison would need revisiting the problem with $I\neq0$ for the axisymmetric case).  \par
{ One could rightly argue that the sideband resonances will change with a different choice of geometry (in this case, a different $\phi$ dependence of $\eta$ and $d$). However, we should keep in mind that the choice is actually not free. In the complete view of the problem, the geometry is an integral part of $\mathbf{B}_0$, and must also satisfy the QS requirements.\cite{jorge2020} A more detailed investigation on the interplay of these geometric requirements is left to future work.  \par
Although successful in localising the PS current singularity, QS seems to inherit from general 3D configurations important non-local effects that complicate the construction of quasisymmetric equilibria. The delicate balance needed to eliminate these sidebands also evidences the sensitive nature of the construction. Any misadjustment could potentially lead to the appearence of islands. \par
} \par
{In Sec.~IVB we indicated that a QS consistent symmetry-resonant surface required $B$ to be a flux function at that surface. From the asymmetric geometry analysis in this section we see that the side-band complication associated to the QS field could also be avoided by making \textit{the entire plasma isodynamic}. { Such a design would reduce PS currents to a minimum, and thus simplify the outset of the problem. Of course, designing a truly isodynamic solution is even more constraining than QS. However, quasi-isodynamic solutions present themselves attractive from this point of view.} These side-band effects presented will also be present in general, non-QS stellarators, but their importance will generally be shadowed by the multiple directly-resonant contributions of the multiple PS current singularities, or additional coupling effects\cite{reiman1986}.} \par
    \begin{figure}
        \centering
        \includegraphics[width=0.5\textwidth]{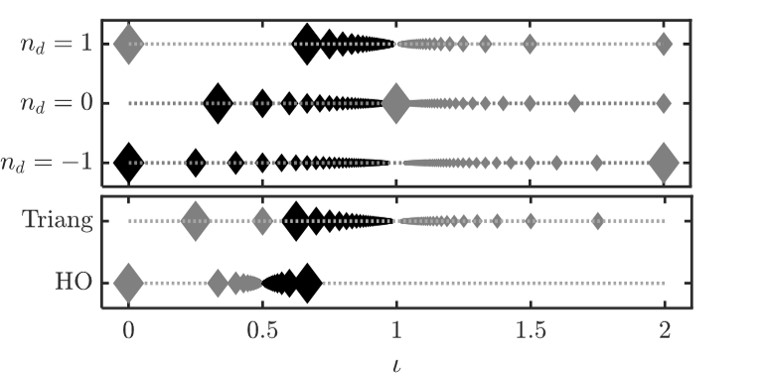}
        \caption{\textbf{Rotational transform line showing coupled resonant surfaces.} Real line showing the coupled rational surfaces to the QS one $\tilde{\alpha}=1$. The top panel shows relevant surfaces coupled to $\tilde{\alpha}=1$ for different $m$ (poloidal mode number and inverse of the size of the symbol), both the $(m\pm n_d,m\pm 2)$ (grey and black) and for three different values of $n_d=-1,0,1$. The bottom panel shows the implications of triangularity for $n_d=1$ and higher order shaping effects for a single mode $m=1$ and $n_d=1$.}
        \label{fig:coupEllipRat}
    \end{figure}

\section{$\delta$-function current singularities and quasisymmetry}
So far in the treatment of current singularities we have only considered the Pfirsch-Schl\"{u}ter piece. We now turn to an analysis of $\delta$-function current singularities in Eq.~(\ref{eqn:currSingEq}). {  The first thing we need to do is to find the amplitude of the $\delta$-function singularities. Once this is done, the associated island size could be estimated by finding the resultant resonant field, following, for example, a procedure similar to that in Sec.~IV. This computation should in principle include all currents, including the PS contribution considered in Sec.~IV in addition to the $\delta$-function singularity. This procedure, in the presence of full geometry, is complex and would mix both effects. This is especially so given that the current singularity depends on the resonant field implicitly. Thus we focus on the $\delta$-function singularity contribution and separate it from the PS-singularity contribution. We shall do so in the simplest of geometries to estimate the contribution it would have to the opening of islands. At the end of this section we shall comment briefly on how the pieces in Sec.~IV and the contribution from the $\delta$-function current singularity would come together. \par
To obtain the amplitude of the $\delta$-function current singularity we need to resolve the singularities solving a boundary layer problem.\cite{hegna1989,cary1983} As in [\onlinecite{hegna1989}], the problem has two parts. One part corresponds to the outer region away from the singularity, where the nested, MHS problem is solved without dealing with the resonant surface explicitly. This solution is generally discontinuous at the resonant surface, leading to a current singularity. The second part of the problem is the inner region, a narrow boundary layer around the resonant surface where the current singularity is resolved. Inside the boundary layer, one allows for the magnetic structure of a reconnected magnetic island and solves the appropriate equilibrium problem there. The size of the current singularity then comes from a balance of the solution outside and inside, and depends explicitly on the size of the island (or equivalently, the size of the resonant field). To find an expression for the potential island then, the resonant field from the computed current is calculated, and the resulting implicit equation solved\cite{bhattacharjee1995}. \par
The amplitude of the current singularity proportional to the island width is also} related to the resistive stability parameter $D_R$.\cite{glasser1976,hegna1989,bhattacharjee1995} To reach this conclusion, one needs to assume a large aspect ratio, small island width, small plasma $\beta$ and neglectible toroidal current (see [\onlinecite{hegna1989}]) in the boundary-layer problem. The consistent island width associated to the $\delta$-function current singularities can then be written in cylindrical geometry as,\cite{bhattacharjee1995}
\begin{equation}
    W_i\approx\frac{r}{m}D_R. \label{eqn:islWidthDr}
\end{equation}
 \par
To understand the implications of QS on this piece of the problem, it is then important to study the form of the resistive parameter.

\subsection{Resistive parameter and quasisymmetry} The resistive parameter $D_R$ was originally obtained by Glasser, Greene and Johnson\cite{glasser1976} in the context of resistive interchange stability. The parameter can be expressed in a variety of equivalent forms. For assessing the differences between the quasisymmetric field and a general configuration it is convenient to write the dependence of $D_R$ on the Jacobian and geometry explicitly. Rewriting the form of $D_R$ in [\onlinecite{glasser1975}] (see Appendix D for more details),
\begin{multline}
    D_R=\left[-\frac{p'\mathcal{J}'_0}{\iota'^2}\Bar{\mathcal{G}}+\left(\frac{p'}{\iota'}\right)^2\mathcal{J}_{0}\bar{\mathcal{G}}\left\langle\frac{1}{B^2}\right\rangle\right]+\\
    +\left(\frac{p'}{\iota'}\right)^2\Bar{\mathcal{G}}\overline{\mathcal{G}\tilde{\mathcal{J}}^{+2}}, \label{eqn:Dr}
\end{multline}
where the modified Jacobian $\mathcal{J}^+=\partial_\alpha\int\mathrm{d}\phi\mathcal{J}$ and $\mathcal{J}_0$ is the 0th harmonic of the Jacobian. The function $\mathcal{G}$ represents the geometry of the configuration through $\mathcal{G}=G/|\nabla\psi|^2$. The overbar operation extracts the resonant component; formally, $\Bar{f}=\oint f\mathrm{d}\phi/\oint\mathrm{d}\phi$ at constant $\alpha$. The tilde functions represent the remaining non-resonant components $\Tilde{f}=f-\Bar{f}$. The expression in Eq.~(\ref{eqn:Dr}) is close to the form by [\onlinecite{hegna1989}].\footnote{There is a typo in [\onlinecite{hegna1989}], Eq.~(51), in the expression for the resistive stability combination $E+F$. There is an additional $\bar{G}$ factor in front of the second term that should not be there.} \par
The first thing to note from the expression for $D_R$ is that, unlike the case of the singular PS current, there is a mixture of terms of very different types. Equation~(\ref{eqn:Dr}) contains a piece that does not depend on the harmonic information of the Jacobian (term in square brackets) and one that does. \par
Let us start from the first piece. This contribution is related to the magnetic well property of the configuration. In fact, the first term is precisely the magnetic well and the second a higher plasma-$\beta$ correction to it. QS does not modify the behaviour of this term in any qualitative way. This is in part because QS does not have an immediate consequence for the properties of the configuration such as $\mathcal{J}_0$, $p'$ and $\iota'$. We thus conclude that this piece will have to be considered case by case, depending on the choice of profiles. This is not to say that the magnetic well piece is unimportant. It is fundamental, as it is the only potentially negative contribution to Eq.~(\ref{eqn:Dr}). From Eq.~(\ref{eqn:islWidthDr}), a negative $D_R$ implies that no island will open. {  The more negative the parameter, the more likely it is that other resonant fields (such as those of Sec. IV) would be shielded.}\par
Thus, any qualitative difference introduced by QS must belong to the destabilising piece in Eq.~(\ref{eqn:Dr}). This is the term involving the modified Jacobian $\mathcal{J}^+$. Let us see in more detail how the differences in the Jacobian affect the function $\mathcal{J}^+$. For a general 3D configuration,
\begin{equation}
    \tilde{\mathcal{J}}^+={\sum}'\frac{\mathcal{J}_{nm}}{\iota_r-n/m}e^{i(m\theta-n\phi)}. \label{eq:jacTild}
\end{equation}
The $'$ means that the exact resonance is not counted when $\iota=n/m$. In the special case of QS, this function takes the simple form,
\begin{equation}
    \Tilde{\mathcal{J}}^+=
    \begin{cases} 
    0 & \iota_r=\tilde{\alpha} \\
    \tilde{\mathcal{J}}\frac{1}{\iota_r-\Tilde{\alpha}} & \iota_r\neq\tilde{\alpha}
 \end{cases}. \label{eq:jacTildQS}
\end{equation}
The function has a single resonance at the quasisymmetric resonant surface. When the symmetry is lost, the resonances appear at many rational values for which $\mathcal{J}_{nm}\neq0$. This simplification is reminescent of the PS situation.
\par
To explore a bit further the structure of the coupling term in Eq.~(\ref{eqn:Dr}), let us explore as way of illustration a term of the form $\overline{\tilde{\mathcal{J}}^+\mathcal{G}}$, asssuming that $\tilde{\mathcal{J}}\sim\exp[im\alpha+im(\iota-\tilde{\alpha})\phi]$. This is actually a term that appears in the coefficient $H$ defining $D_R$ (see Eq.~(\ref{eqn:H})). For this term not to vanish, the geometric piece $\mathcal{G}$ must have matching $\phi$ harmonics at constant $\alpha$. Given the simplicity of $\mathcal{J}^+$ only the $\left(n,m\left[\iota(1-n/m)-\tilde{\alpha}\right]\right)$ Fourier modes of $\mathcal{G}$, where $n,m\in\mathbb{Z}$, will contribute. For the mode to be rational, it must be true that either $n=m$ or $\iota=k/(m-n)$ where $k\in\mathbb{Z}$. The former case corresponds to picking the harmonics matching the QS helicity, while the latter picks side-bands at $m_s(0,\iota-\tilde{\alpha})$, where $m_s$ is an integer $\iota=n_s/m_s$. Thus the barring operation with the Jacobian acts to select very particular geometry modes around the QS resonance. \par
In the general non-QS case, the role that the QS-resonant surface plays is played by many rational surfaces. One thus expects to find an increasingly effective coupling to geometric modes, as well as a multitude of resonances. The presence of a number of resonant points makes it difficult to control $D_R$ from a design point of view. But it also worsens the island opening problem. Recall that any addition from the coupling term is positive. Thus the presence of more resonant pieces worsens the island opening consequences of $\delta$-function current singularities, all other properties kept unchanged.   \par
In practice, even in the general configuration one does not need to consider all the rational surfaces that appear in Eq.~(\ref{eq:jacTild}). This is because in determining the final value of a contribution of a given mode $(n,m)$ to $D_R$, there is a competition between the resonant denominator becoming small, and the corresponding Jacobian mode also becoming smaller for larger mode numbers. Number theory helps us here to understand through a Diophantine approximation how large the resonant contribution is. A given rational number $a/b$ (where $a,b$ are coprime) can be approximated by another rational number of order $m$ no better than by $\sim1/mb$, for $m$ and $b$ coprime. This weak scaling of the resonance is eventually dominated by the exponential decay of the higher-$m$ mode part of the Jacobian. Thus, in practice, only a finite number of lower order rationals are relevant. But even then, the simplicity of QS is apparent. In Fig.~\ref{fig:ratLine} some low-order rationals on the real rotational transform line are represented. \par
\begin{figure}
    \centering \hspace*{-0.5cm}
    \includegraphics[width=0.5\textwidth]{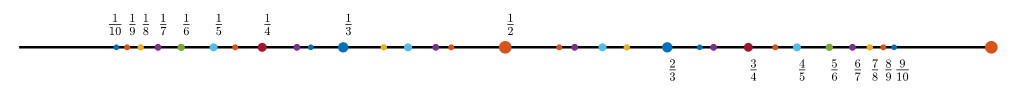}
    \caption{\textbf{A few rational numbers in the real line.} Lowest order rational numbers on the real line in the interval $(0,1]$. The different colors correspond to a different mode number $m$, while the size of the dots represent the importance of the rationals (size in matlab $\propto1/m$).}
    \label{fig:ratLine}
\end{figure}
From what has been presented, it is clear that QS simplifies and reduces these destabilising contributions if other things are kept unchanged. However, whether these differences have practical consequences seem to depend on details of each particular case. \par
{  As indicated at the beginning of this section, these QS effects and those in Sec.~IV are only artificially separated. In reality, the two flavours of singularity occur simultaneously.\cite{hegna1989,bhattacharjee1995} This is important, as the $\delta$-function singularities could reduce some of the side-band effects from the PS contributions we had seen in Sec.~IV. To assess that interaction, the problem in Sec.~IVC would have to be solved including these two singularities together and including the geometric effects. To do so rigorously is complex (and to do it with a self-consistent geomety only possibly attempted numerically) and we do not attempt it here. However, as a first-order approximation one could compare the estimated magnitudes for the island widths and give $W_\mathrm{eff}\sim rD_R/2m+\sqrt{(rD_R/2m)^2+(W_i^{PS})^2}$, where $W_i^{PS}$ is the island estimation from Sec.~IV. For the near-circular geometry and how to put these contributions together we refer to [\onlinecite{bhattacharjee1995}].}

\subsection{A near-axis consideration} Given the competing effects in $D_R$, it is illustrative to consider some particular cases. The near-axis expansion framework\cite{garrenboozer1991a,landreman2018a,rodriguez2020i} provides us with a set of simplified approximate configurations. This permits a more detailed evaluation of the parameter $D_R$. We do not reproduce the expansion here, but point to Eqs.~(4.12) and (4.15) of [\onlinecite{landreman2020}]. These express the parameter $D_R$ for a quasisymmetric and asymmetric configurations to leading order in the aspect ratio $\epsilon$. \par
The leading order form of the resistive stability parameter is $D_R\sim O(\epsilon^{-2})$ in both cases. The quasisymmetric expression includes simpler forms for the expressions in the general case, which in places include toroidal integrals. Otherwise, the expressions are very similar. This similarity in the form of $D_R$ may seem, upon first glance, to be different from the observations made above. \par
Let us see how this is resolved. If we restrict ourselves to the first term in Eq.~(\ref{eqn:Dr}), the magnetic well term, we do expect such similarity. So there is no problem there. The differences detailed above correspond to the destabilising term. In the near-axis expansion framework the difference disappears. This is because to leading order only $m=1$ components of the Jacobian are included. (The $m$-th Fourier component of the Jacobian scales like $\epsilon^m$.) Then, the leading contribution does not allow the general configuration to pick on its multiple contributing resonant denominators. This makes both forms for $D_R$ very similar to one another. \par
For the differences to become significant, higher orders in the expansion are needed. For those stellarator equilibria designed based on near-axis procedures, such as those presented in [\onlinecite{landreman2020}], the higher order terms in $D_R$ may become important when the coefficient $D_R$ calculated using the leading order near-axis expansion significantly deviates from the full equilibrium design. We expect to find these higher order effects more relevant further away from the axis and wherever higher $m$ components of the Jacobian are significant. From this analysis we may take that the behaviour close to the magnetic-axis requires case-by-case consideration, with QS having a reduced effect on $D_R$. \par
Before concluding, we enphasise the importance of the location and magnitude of resonances. We saw the selective nature of the PS singularities in QS configurations, as well as the changes in the magnitude of $\delta$-function current singularities. Even if each island has been treated separatedly, if they lie close to one another, they may overlap. If they do so, stochastic regions generally appear in the problem severely affecting confinement. This may be related to $\beta$ limits\cite{nakamura1990,loizu2017}, which would definitely be affected by QS.


\section{Conclusion}
A study of current singularities { and potential magnetic resonances} in a quasisymmetric magnetic field with nested flux surfaces was performed. Although a strictly quasisymmetric field was shown incompatible with multiple magnetic island chains, the presence of current singularities and Pfirsch-Schl\"{u}ter currents was explored. The appearence of islands when singularities are relaxed was investigated in the context of `mostly' quasisymmetric configurations. \par
Largely as a consequence of the simple harmonic content of $B$, significant differences with the behaviour of general 3D magnetic configurations are observed. Generally, this QS property confers a resilience to Pfirsch-Schl\"{u}ter singular currents. To leading order they may only appear on the magnetic surface resonant with the symmetry helicity. Leaving this surface outside the configuration thus avoids the opening of direct-resonant islands. We perturbatively showed that the PS currents, even in a mild ellipticity of surfaces lead to finite resonant fields at other rationals, unlike an axisymmetric configuration. { This intricate 3D structure suggests the complexity in constructing an exactly QS MHS equilbrium. Codes that are based on assumption of nested flux surfaces everywhere remain vulnerable to island formation unless the optimization process builds in some of the criteria discussed in this paper.} \par
When considering the presence of $\delta$-function singularities, the differences with the general 3D configuration are more subtle. The opening of magnetic islands is controlled by the parameter $D_R$ as shown in [\onlinecite{hegna1989}], to which there are two main contributions. In scenarios where $D_R$ is not dominated by the magnetic well term and higher modes of $B$ are relevant, a general stellarator presents the possibility of increased destabilisation from lower-order rationals. This is avoided in the quasisymmetric scenario in a similar way to the PS current behaviour. Whether those terms avoided are relevant contributors to $D_R$ is a case-by-case question. \par
{  These two contributions have been investigated separatedly, but they occur simultaneously.\cite{hegna1989,bhattacharjee1995,cary1985} An analysis that considers both at the same time taking the full geometry into account is not presented here, but an estimate has been suggested at the end of Sec. VA.} Studies of these fundamental differences in practical realisations of the quasisymmetric fields with a self-consistent geometry will probably need numerical approaches, left for future work.


\appendix
\section{Mercier and Boozer coordinates for elliptical surfaces}
In this appendix we address the construction of the coordinate mapping between the cylindircal Mercier coordinates and Boozer flux coordinates close to the magnetic axis. To complete this comparison, we start by describing Mercier coordinates\cite{Helander2014}, which gives positions in space according to
\begin{equation}
    \n{x}=\n{X}(l)+\hat{\kappa}(l)\rho\cos\vartheta+\hat{\tau}(l)\rho\sin\vartheta, \label{eqn:xMercDef}
\end{equation}
and 
\begin{equation}
    \vartheta=\omega-\int_0^l\tau\mathrm{d}l'. \label{eqn:omegThetDef}
\end{equation}
Here the Frenet vectors are used as a basis for $\n{x}$, with $\n{X}$ representing the magnetic axis and $l$ parametrising its length. From the definition, $\vartheta$ and $\rho$ are cylindrical coordinates in the plane perpendicular to the axis at each $l$; the coordinate $\omega$ is simply a redefinition through Eq.~(\ref{eqn:omegThetDef}) of the cylindrical angle (redefining the zero at each plane). \par
We commence the comparison to Boozer coordinates with coordinate $l$, which parameterises the length along the magnetic axis. From near-axis expansions of quasisymmetric fields\cite{garrenboozer1991b,landreman2018a,rodriguez2020i} it holds that $\mathrm{d}l/\mathrm{d}\phi=\mathrm{const.}$ That is, the Boozer toroidal angle is a scaled version of the axis length parametrisation $l=L\phi/2\pi$, where $L$ is the total length of the magnetic axis. \par
The coordinate relation is not that simple for the other two coordinates. To aid with the comparison, it is important to find a way to represent the magnetic field in Mercier coordinates,
\begin{equation}
    \n{B}=[B_{l0}+O(\rho)]\nabla l+[B_{\omega1}\rho^2+O(\rho^2)]\nabla\omega+[B_{\rho1}\rho+O(\rho^2)]\nabla\rho, \label{eq:magAxApprox}
\end{equation}
where due to quasisymmetric considerations $B_{l0}$ is a constant and the corresponding expansions are a consequence of the magnetic axis, of course, being parallel to the magnetic field. Comparing this form to the Boozer form in Eq.~(\ref{eqn:fieldCoCon}) will then provide important information to complete the coordinate maping. For once, the coefficient $B_{l0}$ can be rewritten as $B_{l0}\approx GL/2\pi$. \par
An additional important piece for the coordinate mapping is the elliptical shape of the magnetic surface cross-sections. Define an ellipse in a constant $l$ plane as \cite{Helander2014},
\begin{align*}
    x = re^{-\eta/2}\cos v \\
    y = re^{\eta/2}\sin v,
\end{align*}
where the major axis of the ellipse lies along $y$. Comparing this to Eq.~(\ref{eqn:xMercDef}), and defining the angle between the curvature vector and the minor axis to be $d(l)$, 
\begin{align}
    \tan u&=e^\eta \tan v, \\
    r^2&=\rho^2\left(e^\eta\cos^2u+e^{-\eta}\sin^2 u\right), \label{eqn:r2}
\end{align}
where $\omega=u-d(l)+\int_0^l\tau\mathrm{d}l'$. \par
The expression in Eq.~(\ref{eqn:r2}) can be immediatly related to the area of the elliptical cross-sections, and thus the magnetic flux coordinate $\psi$. It follows that,
\begin{equation*}
    2\pi\psi\approx B_{l0}\pi r^2= B_{l0}\pi\rho^2(e^\eta\cos^2 u+e^{-\eta}\sin^2 u).
\end{equation*}
At this point, the only coordinate left for comparison is the angular `poloidal' $\omega$. To advance, we shall look at the contravariant form of the magnetic field in Eq.~(\ref{eqn:fieldCoCon}), and compare it to Eq.~(\ref{eq:magAxApprox}). Interpreting the equation as one for $\alpha$ in terms of Mercier coordinates, to leading order,\footnote{We are taking $\partial_\rho \alpha/\partial_\omega\alpha$ not to be $O(1/\rho)$. This allows us to drop higher order terms in $\rho$.}
\begin{align*}
    \partial_\omega \alpha\approx\frac{\rho B_{l0}}{\psi_\rho}&=\frac{1}{\cosh\eta+\sinh\eta\cos2u}.
\end{align*}
Exactly integrated,
\begin{equation}
    \tan\bar{\alpha}=\tan[\alpha+C(l)]=e^{-\eta}\tan u. \label{eq:alpuOrd1}
\end{equation}
In the leading $O(\rho)$ part of the equation equating Eq.~(\ref{eqn:fieldCoCon}) and Eq.~(\ref{eq:magAxApprox}) there is not enough information to uniquely determine the form of $\alpha$. To complete the construction one should resort to higher orders. In particular, we have two additional magnetic field components along $\nabla\omega$ and $\nabla\rho$, so that
\begin{align}
    \partial_\omega\psi\partial_l\alpha-\partial_l\psi\partial_\omega\alpha=B_{1\rho}\rho^2 \label{eq:alpOme1}\\
    -\partial_\rho\psi\partial_l\alpha=B_{\omega1}. \label{eq:alpOme2}
\end{align}
We further need a way to relate the magnetic field components, and we shall use the assumption that there is no net toroidal current associated with the field. Thus, using Amp\`{e}re's law,
\begin{equation*}
    \nabla\times\n{B}_0\cdot\nabla l=0\implies 2B_{\omega1}=\partial_\omega B_{\rho1}.
\end{equation*}
Using this latter relation, and putting together equations (\ref{eq:alpOme1}) and (\ref{eq:alpOme2}),
\begin{equation*}
    (\partial^2_{\omega}\psi+2\partial_\rho\psi)\partial_l\alpha=\partial_l\psi\partial^2_\omega\alpha+\partial_l\partial_\omega\psi\partial_\omega\alpha-\partial_\omega\psi\partial_l\partial_\omega\alpha.
\end{equation*}
Using the form of $\psi$ associated to our elliptical flux surface approximation and the expression for $\partial_\omega\alpha$, we may write,
\begin{equation*}
    \partial_l\alpha=\frac{\partial_l\partial_\omega\psi}{2\partial_\rho\psi\cosh\eta}.
\end{equation*}
To obtain an expression for $C(l)$, using Eq.~(\ref{eq:alpuOrd1}),
\begin{align*}
    C_l=-\alpha_l+\frac{1}{\psi_\rho}(d'-\tau-\frac{1}{2}\eta'\sin2u)=\frac{d'-\tau}{\cosh\eta},
\end{align*}
which upon integration (and up to a constant) yields,
\begin{equation}
    C(l)=\int_0^l\frac{d'-\tau}{\cosh\eta}\mathrm{d}l'.
\end{equation}
Putting it all together, it follows that
\begin{equation}
    \tan\left[\omega+d-\int_0^l\tau\mathrm{d}l'\right]=e^\eta\tan\left[\alpha+\int_0^l\frac{d'-\tau}{\cosh\eta}\mathrm{d}l'\right]. \label{eq:alpOmegMerc}
\end{equation}
This concludes the coordinate mapping needed to leading order. 

\section{Deriving the coupled side-band equations}
{  In this Appendix we detail the derivation of the equations in Section IVC. In order to do so, we first need to introduce some important tools. As we will be using Fourier series it is important to know,
\begin{equation}
    \partial_\omega(n\phi-m\theta)=-\frac{m}{e^\eta\cos^2u+e^{-\eta}\sin^2u}, \label{eqn:domegaFour}
\end{equation}
as well as the following,
\begin{equation*}
    \cosh\eta+\sinh\eta\cos 2u=\frac{1}{\cosh\eta-\sinh\eta\cos 2\bar{\alpha}}.
\end{equation*}
With this and the expression for $\psi$ in Mercier coordinates as given in Eq.~(\ref{eqn:MercBoozPsi}) we have most of the tools we need to proceed. \par
Let us then start showing how Eq.~(\ref{eqn:divHEllip}) is obtained; i.e., an expression for $\nabla\cdot\mathbf{H}$. To evaluate the divergence we need the covariant description of $\mathbf{H}$ in Mercier coordinates. From Eq.~(\ref{eqn:Hfield}) it is clear we need, ignoring the $\nabla l$ part, 
\begin{equation*}
    \nabla\psi\approx\frac{2\psi}{\rho}\nabla\rho-B_{l0}\rho^2\sinh\eta\sin2u \nabla\omega.
\end{equation*}
This implies that,
\begin{gather}
    H_\rho=H_\rho^\circ\left(\cosh\eta+\sinh\eta\cos 2u\right), \\
    H_\omega=-H_\rho^\circ\sinh\eta\sin2u. \label{eqn:Homega}
\end{gather}
The second is the component along $\hat{\omega}=\rho\nabla\omega$ of $\mathbf{H}$, in the form that will be needed to compute the divergence of $\mathbf{H}$ ---think of how one would compute the divergence in cylindrical polar coordinates. Here $H_\rho^\circ$ is the circular-geometry expression,
\begin{equation}
    H_\rho^\circ=B_{l0}\rho \sum \frac{p'}{inM(\tilde{\alpha}-\iota)}\mathcal{J}_ne^{in(N\phi-M\theta)}, \label{eqn:HcircApp}
\end{equation}
as in Eq.~(\ref{eqn:Hcirc}). Now, $\nabla\cdot\mathbf{H}\approx \partial_\rho(\rho H_\rho)/\rho+\partial_\omega H_\omega/\rho$, as argued in the main text, Eq.~(\ref{eqn:divH}), dropping $\partial_l$ piece. So we need explicitly these two pieces. The latter is of particular interest, as it is a piece that only occurs when finite ellipticity is considered. Acting with $\partial_\omega$ on the $\sin2u$ factor of Eq.~(\ref{eqn:Hcirc}) and the Fourier exponent of Eq.~(\ref{eqn:HcircApp}),
\begin{equation}
    \partial_\omega H_\omega\approx -2\sinh\eta\cos2u H_\rho^\circ+\frac{B_{l0}\rho\sinh\eta\sin2u}{\cosh\eta+\sinh\eta\cos2u}\frac{p'}{\tilde{\alpha}-\iota}\tilde{\mathcal{J}},
\end{equation}
where once again we have dropped terms that are $O(\eta^2)$ or larger. That is why we have not acted upon the $\psi$ dependent pieces explicitly. This would bring additional geometric factors as can be seen from Eq.~(\ref{eqn:MercBoozPsi}). Then, 
\begin{align}
    \nabla\cdot\mathbf{H}&\approx \frac{1}{\rho}\partial_\rho(\rho H_\rho)+\frac{1}{\rho}\partial_\omega H_\omega\approx\nonumber\\
    &\approx \frac{1}{\rho}\partial_\rho(\rho H_\rho^\circ)+\rho\partial_\rho\left(\frac{H_\rho^\circ}{\rho}\right)\sinh\eta\cos2u+\nonumber\\
    &~~~~~~~~~~~+\frac{B_{l0}p'}{\tilde{\alpha}-\iota}\tilde{\mathcal{J}}\sinh\eta\sin2\bar{\alpha}\nonumber\\
    &\approx \frac{1}{\rho}\partial_\rho(\rho H_\rho^\circ)-\mathcal{H}e^{2i\bar{\alpha}}, \label{eqn:divHapp}
\end{align} 
where
\begin{equation}
    \mathcal{H}=i\frac{B_{l0}p'}{\tilde{\alpha}-\iota}\tilde{\mathcal{J}}\sinh\eta.
\end{equation}
We have here assumed that $\mathcal{J}_n\propto\psi^{m/2}\propto\rho^m$ and $p'/(\tilde{\alpha}-\iota)$ is constant. The latter is a reasonable choice so long as we are away from the resonance, which we generally are when we compute the geometry-coupled off-resonant contributions. Both these assumptions are key for the expression for $\nabla\cdot\mathbf{H}$ to have a single $2\bar{\alpha}$ harmonic contribution and no $-2\bar{\alpha}$. \par
The radial operator in Eq.~(\ref{eqn:divHapp}) can be rewritten in terms of the pseudo-radial coordinate $\bar{\rho}=\sqrt{2\psi/B_{l0}}$,
\begin{equation*}
    \frac{1}{\rho}\partial_\rho(\rho H_{\rho,nm}^\circ)=\frac{1}{\bar{\rho}}\partial_{\bar{\rho}}(\bar{\rho}H_{\bar{\rho},nm}^\circ),
\end{equation*} 
where we have defined the flux function $H_{\rho,nm}^\circ/\rho$ as $H_{\bar{\rho},nm}^\circ=(\bar{\rho}/\rho)H_{\rho,nm}^\circ$, and are focusing on Fourier modes. So, omitting the sum over the Fourier exponents,
\begin{equation}
    \nabla\cdot\mathbf{H}\approx\frac{1}{\bar{\rho}}\partial_{\bar{\rho}}(\bar{\rho}H_{\bar{\rho},nm}^\circ)-\mathcal{H}_{nm}e^{2i\bar{\alpha}}. \label{eqn:divHappFin}
\end{equation}
Let us now focus on the Laplacian piece in Eq.~(\ref{eq:poissRes}). The procedure to follow is similar to the kind of manipulations we have made so far. Let us start by considering the $\partial_\omega$ piece in Eq.~(\ref{eq:laplcianMercer}). Once again, we need to be careful in applying the operator, as we need to act on the Fourier exponential when writing down $F$ in its Fourier series, but also act on the flux dependent part of the expression. The Fourier coefficients $F_{nm}(\psi)$ are flux functions by construction. Thus using the map in Eq.~(\ref{eqn:MercBoozPsi}),
\begin{equation*}
    \partial_\omega F_{nm}=-B_{l0}\rho^2\sinh\eta\sin2u\partial_\psi F_{nm}.
\end{equation*}
Therefore, using Eq.~(\ref{eqn:domegaFour}) and keeping terms to leading order in the shaping $\eta$, 
\begin{align*}
    \partial_\omega^2F\approx \sum_{n,m} e^{i(n\phi-m\theta)}\left[-m^2+e^{2i\bar{\alpha}}\hat{\Xi}_2+e^{-2i\bar{\alpha}}\hat{\Xi}_{-2}\right]F_{nm},
\end{align*}
where to leading order in $\eta$,
\begin{gather}
    \hat{\Xi}_2\approx m(m-1)\sinh\eta+2\psi(m-1)\sinh\eta\partial_\psi, \\
    \hat{\Xi}_{-2}\approx m(m+1)\sinh\eta-2\psi(m+1)\sinh\eta\partial_\psi.
\end{gather}
To arrive here we used the convenient expressions,
\begin{gather*}
    \sin2u=\frac{\sin2\bar{\alpha}}{\cosh\eta-\sinh\eta\cos2\bar{\alpha}}, \\
    \cos2u=\frac{\cosh\eta\cos2\bar{\alpha}-\sinh\eta}{\cosh\eta-\sinh\eta\cos2\bar{\alpha}}.
\end{gather*}
We note that these $\Xi$ expressions constitute operators that act on $\psi$ dependent functions. \par
We should now focus on the radial piece and try to cast it in terms of $\bar{\rho}$ as we did with the radial part of $\nabla\cdot\mathbf{H}$. This is vital to be able to obtain a mode resolved form for the equations, as the Fourier coefficients are functions of $\psi$. After doing the appropriate transformation, the Laplacian we are left with (omitting this time the summation sign and the Fourier exponential) reads
\begin{equation*}
    \nabla^2 F\approx\frac{\bar{\rho}^2}{\rho^2}\frac{1}{\bar{\rho}}\partial_{\bar{\rho}}(\bar{\rho}\partial_{\bar{\rho}} F_{nm})+\frac{1}{\rho^2}\left[-m^2+e^{2i\bar{\alpha}}\hat{\Xi}_2+e^{-2i\bar{\alpha}}\hat{\Xi}_{-2}\right] F _{nm}.
\end{equation*}
Together with the expression obtained in Eq.~(\ref{eqn:divHappFin}), we obtain the following form for Eq.~(\ref{eq:poissRes}),
\begin{multline}
    \frac{\bar{\rho}^2}{\rho^2}\frac{1}{\bar{\rho}}\partial_{\bar{\rho}}(\bar{\rho}\partial_{\bar{\rho}} F_{nm})+\frac{1}{\rho^2}\left[-m^2+e^{2i\bar{\alpha}}\hat{\Xi}_2+e^{-2i\bar{\alpha}}\hat{\Xi}_{-2}\right] F _{nm}\approx \\
    \frac{1}{\bar{\rho}}\partial_{\bar{\rho}}(\bar{\rho}H_{\bar{\rho}}^\circ)-\mathcal{H}e^{2i\bar{\alpha}}.
\end{multline}
We would like to get rid of the $\rho$ pieces so that we have all angle dependent pieces explicitly. Multiplying through $\rho^2/\bar{\rho}^2$ and noting that $\rho^2/\bar{\rho}^2=(\cosh\eta+\sinh\eta\cos2u)^{-1}=\cosh\eta-\sinh\eta\cos2\bar{\alpha}$, to leading order we may write,
\begin{multline}
    \frac{1}{\bar{\rho}}\partial_{\bar{\rho}}(\bar{\rho}\partial_{\bar{\rho}} F_{nm})+\frac{1}{\bar{\rho}^2}\left[-m^2+e^{2i\bar{\alpha}}\hat{\Xi}_2+e^{-2i\bar{\alpha}}\hat{\Xi}_{-2}\right]F _{nm}\approx\\
    \frac{1}{\bar{\rho}}\partial_{\bar{\rho}}(\bar{\rho}H_{\bar{\rho}}^\circ)+\mathcal{H}_2e^{2i\bar{\alpha}}+\mathcal{H}_{-2}e^{-2i\bar{\alpha}},
\end{multline}
where
\begin{gather}
    \mathcal{H}_2\approx -\mathcal{H}-\frac{1}{2}\sinh\eta\frac{1}{\bar{\rho}}\partial_{\bar{\rho}}(\bar{\rho}H_{\bar{\rho}}^\circ), \\
    \mathcal{H}_{-2}\approx -\frac{1}{2}\sinh\eta\frac{1}{\bar{\rho}}\partial_{\bar{\rho}}(\bar{\rho}H_{\bar{\rho}}^\circ).
\end{gather}
This completes the derivation of the side-band equations in Sec.~IVC.}

\section{Island widths for resonant circular and non-resonant elliptical cross-section}
Let us start finding an analytic form for the island width that the singular PS current would give rise to at the quasisymmetric resonant surface without any geometric coupling. This is precisely the situation described in [\onlinecite{reiman1984}], and we refer to this work for more details. The piece of information needed to find an expression for the magnetic island width is $B_{\rho nm}$ at $\iota=n/m$; i.e., the resonant piece of $\n{B}$ normal to the flux surface as it appears in Eq.(\ref{eqn:islWidthB}). To find the relevant form of $B_{\rho nm}$, we need first to solve Eq.~(\ref{eqn:possFourCirc}) for the given form of $H_{\rho nm}^\circ$, to then obtain $B_{\rho nm}=\partial_\rho F_{nm}+H_{\rho nm}$. The solution to Eq.~(\ref{eqn:possFourCirc}) is,
\begin{align}
    F_{nm}=&-\frac{1}{2}\rho^m\int_a^\rho\rho^{-m}H^\circ_{\rho nm}\mathrm{d}\rho-\nonumber\\
    &-\frac{1}{2}\rho^{-m}\int_0^\rho\rho^{m}H^\circ_{\rho nm}\mathrm{d}\rho ~~~~~(0\leq\rho<a) \nonumber\\
    =&-\frac{1}{2}\rho^{-m}\int_0^a\rho^mH^\circ_{\rho nm}\mathrm{d}\rho~~~~~~(\rho\geq a), \label{eqn:Fcirc}
\end{align}
where $a$ denotes the edge of the plasma. Note that the general homogeneous solution to the equation reads $F_{nm}=A\rho^m+B\rho^{-m}$, and the particular form obtained follows from requiring three conditions. First, the physical requirement of asymptotic vanishing of $F$. Second, the continuity of $F_{nm}$ across the plasma edge $\rho=a$. And finally, the regularity condition at the origin, which amounts to $F_{nm}\propto\rho^m$ for $\rho\sim0$. \par
With this form of the solution, we obtain
\begin{multline*}
    B_{\rho nm}=-\frac{m}{2}\left[\rho^{m-1}\int_a^\rho \rho^{-m}H^\circ_{\rho nm}\mathrm{d}\rho-\right.\\
    \left.-\rho^{-(m+1)}\int_0^\rho\rho^mH^\circ_{\rho nm}\mathrm{d}\rho\right].
\end{multline*}
Thus all there is to be done is to find a closed form for the integrals of $H^\circ_{\rho nm}$. To do so, and because we are for simplicity considering the region of the plasma close to the magnetic axis in the near-axis sense, we need to estimate the behaviour of some of the functions in $H^\circ_{\rho nm}$. Start with the rotational transform; because we are looking at the resonant contribution, the rotational transform shear comes into play with $\Tilde{\alpha}-\iota=-(\rho-\rho_r)\iota'$, where $\rho_r$ is the rational surface location. Further, take the pressure to be modelled as $p=p_0(1-\rho^2/a^2)$ and from regularity on axis, the leading order form of $\mathcal{J}_{nm}\propto\rho^m$. We need to be careful with $p'$, which denotes a flux derivative and thus may be written as $p'=-2p_0/B_{l0}a^2$. With all this arsenal, we write
\begin{equation*}
    H_{\rho n}\approx \rho\frac{2p_0/a^2}{(\rho-\rho_r)\iota'}\frac{\rho^{nM}}{inM}\hat{\mathcal{J}}_{n},
\end{equation*}
where $m=nM$. Thus,
\begin{multline}
    B_{1\rho nm}\approx\frac{p_0}{2i\iota' a^2}\mathcal{J}_n\left\{\frac{a}{\rho_r}+\ln\left(\frac{a-\rho_r}{\rho_r}\right)+\sum_{k=2}^{2m+1}\frac{1}{k}\right\}, \label{eqn:B1rCirc}
\end{multline}
for which it was convenient to write $\rho^{2m+1}/(\rho-\rho_r)=\sum_{k=0}^{2m}\rho^{2m-k}\rho_0^k+\rho_0^{2m+1}/(\rho-\rho_0)$. In performing these integrals logarithmic divergences cancel out from the two pieces of the integral in $B_{\rho nm}$, and the divergence at $\rho=\rho_0$ is dropped assuming $p$ to be flat there. This is a usual assumption. The estimated island width at the resonant QS surface may then be explicitly written using Eq.~(\ref{eqn:B1rCirc}) into Eq.~(\ref{eqn:islWidthB}).\cite{reiman1984} \par
To deal with the elliptical geometry situation a similar problem to the one here sketched needs to be solved. We are here going to find the resonant magnetic field piece associated to the $(n+n_d,m+2)$ mode, which obeys to leading order Eq.~(\ref{eqn:poissEllip}). The solution to this equation is in form very similar to Eq.~(\ref{eqn:possFourCirc}), and so, restricting $m>0$,
\begin{multline}
    F_{n+n_d,m+2}\approx\frac{1}{2(m+2)}{\bar{\rho}}^{m+2}\int_{\bar{\rho}}^\infty\Lambda_{-2}\zeta^{-(m+3)}\mathrm{d}\zeta+\\
    +\frac{1}{2(m+2)}{\bar{\rho}}^{-(m+2)}\int_0^{\bar{\rho}}\Lambda_{-2}\zeta^{m+1}\mathrm{d}\zeta,
    \label{eqn:Fside}
\end{multline}
where we have defined $\Lambda_{-2}=\hat{\Xi}_{-2}F_{nm}-\bar{\rho}^2\mathcal{H}_{-2}$ for the inhomogeneous part of Eq.~(\ref{eqn:poissEllip}). One might be tempted to use for $F_{nm}$ the expression we calculated for the circular resonant case. However that would not be right, as the expression we found was for the value of $F_{nm}$ at the resonance. Here we are not at resonance, and thus the integrals in Eq.~(\ref{eqn:Fcirc}) need to be reevaluated. In doing so we assume now $\tilde{\alpha}-\iota$ to be roughly a constant (rather than it going through zero). As a result, 
\begin{align}
    F_{nm}&=\frac{p_0}{2im(\tilde{\alpha}-\iota)}\mathcal{J}_n\left(\frac{m+2}{m+1}\frac{{\bar{\rho}}^2}{a^2}-1\right) ~~~~~(0\leq{\bar{\rho}}<a) \nonumber\\ 
    &=\frac{p_0}{2im(m+1)(\tilde{\alpha}-\iota)}\mathcal{J}_n(a)\left(\frac{{\bar{\rho}}}{a}\right)^{-m} ~~~~~({\bar{\rho}}>a).\label{eqn:nonResF}
\end{align}
Then, acting on it with $\hat{\Xi}_{-2}$,
\begin{equation*}
    \frac{1}{\bar{\rho}^2}\hat{\Xi}_{-2}F_{nm}\approx-\frac{1}{\bar{\rho}^2}\sinh\eta\frac{p_0}{im(\tilde{\alpha}-\iota)}
    \begin{cases} 
        \mathcal{J}_n(m+2)\frac{{\bar{\rho}}^2}{a^2} & 0\leq {\bar{\rho}}<a \\
        -m\mathcal{J}_n(a)\left(\frac{a}{{\bar{\rho}}}\right)^m & {\bar{\rho}}\geq a
     \end{cases}.
\end{equation*}
The piece $\mathcal{H}_{-2}$,
\begin{equation*}
    \mathcal{H}_{-2}\approx \frac{p_0}{ima^2}\frac{\mathcal{J}_n}{\tilde{\alpha}-\iota}(m+2)\sinh\eta.
\end{equation*}
These pieces can then be summed to form $\Lambda_{-2}$ and directly find Eq.~(\ref{eqn:Fside}). \par
Of course what we are after is not just $F$, but rather the resonant component of $\mathbf{B}=\mathbf{H}+\nabla F$. It is the resonant component normal to the flux surface that gives rise to the potential opening of islands. In terms of the variables we have, the component along $\bar{\rho}$, our pseudo-radial coordinate, is the relevant component. In fact, the one that would appear in Eq.~(\ref{eqn:islWidthB}). \par
The contribution from $\nabla F$ to this component $\nabla F\cdot\nabla\bar{\rho}\approx\partial_{\bar{\rho}}F$, but not quite equal. This is a result of Boozer coordinates not corresponding to an orthogonal coordinate system. That is, in computing the projection along the flux surface normal, one would in principle also have $\partial_\theta F\nabla\theta\cdot\nabla\bar{\rho}$ and similarly for $\phi$. However, we may say something more precise about these by using the mapping of Boozer to Mercier coordinates --the latter being an orthogonal basis. To leading order in the shaping $\eta$, only the $\partial_{\bar{\rho}}$ piece is important as Eqs.~(\ref{eqn:MercBoozPhiC})-(\ref{eqn:MercBoozPsiC}) hold. Any higher shaping will give rise to higher order geometric corrections. Thus, we just have to concentrate on $\partial_{\bar{\rho}}F_{n+n_d,m+2}$, which yields 
\begin{multline}
    \partial_{\bar{\rho}} F_{n+n_d,m+2}\approx \\ 
    \frac{p_0}{4im(\tilde{\alpha}-\iota)}\sinh\eta\frac{\mathcal{J}_n}{\bar{\rho}}\left[\frac{3m+2}{m+1}-4(m+2)\ln\left(\frac{a}{\bar{\rho}}\right)\right]\frac{\bar{\rho}^2}{a^2}.
\end{multline}
This is the contribution to the resonant field from $F$. There could also be a piece coming from $\mathbf{H}$ directly. In this case, we are again interested in looking at the projection $\mathbf{H}\cdot\nabla\bar{\rho}$ that resonates with the $(n+n_d,m+2)$ mode. The $\nabla\psi$ term from Eq.~(\ref{eqn:Hfield}) does not have a piece resonant with this side-band mode, and thus it does not contribute to the resonant field. The other possible contribution coming from the $\nabla\phi$ piece can be evaluated going once again to Mercier coordinates. In this case keeping geometric effects in $\eta$ to leading order,
$$ \nabla\phi\cdot\nabla(\bar{\rho}^2)\approx -(4\pi/L)\sinh\eta(d'-\tau)\sin2\bar{\alpha}. $$ 
So there is potential non-zero coupling in this term that could contribute to the $(n+n_d,m+2)$ mode. However, the expression accompanying $\nabla \phi$ is only a flux function. This means that this piece will have a main mode $\pm(n_d,2)$ with additional modes coming from the toroidal dependence of $d$ and $\tau$ (and possibly $\eta$). The lack of an explicit relation to the symmetry $(n,m)$ mode makes us generally take it to have no net contribution to the resonant $B_{1\rho}$ at the $\iota=(n+n_d)/(m+2)$ surface. Then,
\begin{multline}
    B_{\rho,n+n_d,m+2}\approx \frac{p_0}{4im(\tilde{\alpha}-\iota)}\sinh\eta\frac{\mathcal{J}_n}{\bar{\rho}}\left[\frac{3m+2}{m+1}-\right.\\
    \left.-4(m+2)\ln\left(\frac{a}{\bar{\rho}}\right)\right]\frac{\bar{\rho}^2}{a^2}.
\end{multline}
This radial field will open a magnetic island of size as given in Eq.~(\ref{eqn:islWidthB}) not at the quasisymmetric resonant surface but rather at $\iota=(n+n_d)/(m+2)$. 

\section{Derivation of Jacobian form of $D_R$}
Let us start by invoklng the forms of the pieces that make $D_R$ up, from [\onlinecite{glasser1975}],
\begin{align}
    E=\frac{1}{\Lambda^2(V')^2}\left\langle\frac{B^2}{|\nabla \psi|^2}\right\rangle\left[-p'\frac{V''}{(V')^3}\right.&\left.+\Lambda\frac{\langle\sigma B^2\rangle}{\langle B^2\rangle}\right] \label{eqn:E}\\
    F=\frac{1}{\Lambda^2(V')^4}\left\langle\frac{B^2}{|\nabla \psi|^2}\right\rangle\left[\left\langle\frac{\sigma^2 B^2}{|\nabla\psi|^2}\right\rangle-\right.&\left.\frac{\langle\sigma B^2/|\nabla\psi|^2\rangle^2}{\langle B^2/|\nabla\psi|^2\rangle}+\right.\nonumber\\
    &\left.+(p')^2\left\langle\frac{1}{B^2}\right\rangle\right], \label{eqn:F} \\
    H=\frac{1}{\Lambda(V')^2}\left\langle\frac{B^2}{|\nabla \psi|^2}\right\rangle\left[\frac{\langle\sigma B^2/|\nabla\psi|^2\rangle}{\langle B^2/|\nabla\psi|^2\rangle}\right.&\left.-\frac{\langle\sigma B^2\rangle}{\langle B^2\rangle}\right],  \label{eqn:H}
\end{align}
where $V$ is the volume enclosed by constant $\psi$ surfaces, $\sigma=\n{j}\cdot\n{B}/B^2$, primes denote derivatives with respect to $\psi$ and $\Lambda=4\pi^2\iota'/(V')^3$ (where the factors of $2\pi$ come from the use of angular coordinates normalised to 1 rather than $2\pi$ in the original work by [\onlinecite{glasser1975}]). Then $D_R=E+F+H^2$ by definition. We need to further specify the meaning of the angled brackets in the expressions Eqs.~(\ref{eqn:E})-(\ref{eqn:H}), 
\begin{equation*}
    \langle f\rangle=\frac{\oint\frac{f\mathrm{d}l}{B}}{\oint\frac{\mathrm{d}l}{B}}=\frac{\oint\frac{f\mathrm{d}\phi}{B^2}}{\oint \frac{\mathrm{d}\phi}{B^2}}.
\end{equation*}
This average along closed field lines may be related to a flux surface average by application of Weyl's lemma\cite{Helander2014} when high order rationals are involved. We shall assume this here for simplicity. With this definition, the expression may be related to the  bar operation that picks the resonant component of the operated function,\cite{glasser1975}
\begin{equation*}
    \Bar{f}=\frac{\oint f\mathrm{d}\phi}{\oint\mathrm{d}\phi}=\frac{\langle fB^2\rangle}{\langle B^2\rangle}.
\end{equation*}
It is convenient to define $\mathcal{G}=G/|\nabla\psi|^2$ as a function that represents the geometry aspects of the configuration. Of course, to proceed further, we need to find closed form expressions for functions such as $\sigma$ in terms of configuration properties. Using Eq.~(\ref{eqn:fieldCoCon}) and the MHS balance $\n{j}\times\n{B}=\nabla p$, it follows that (with $\alpha$ the field line label),
\begin{equation*}
    \partial_\phi B_\psi|_\alpha=\mathcal{J}p'+G',
\end{equation*}
from which,
\begin{equation}
    \sigma=-p'\tilde{\mathcal{J}}^+.
\end{equation}
With this, we may proceed to rewrite Eqs.~(\ref{eqn:E}) and (\ref{eqn:F}) first in the form,
\begin{equation}
    E=\frac{1}{\Lambda^2(V')^2}\overline{\mathcal{G}}\frac{\langle B^2\rangle}{G}\left(-p'\frac{V''}{(V')^3}\right)
\end{equation}
\begin{multline}
     F=\frac{1}{\Lambda^2(V')^4}\overline{\mathcal{G}}\frac{\langle B^2\rangle^2}{G^2}\left[\overline{\tilde{\mathcal{J}}^{+2}\mathcal{G}}-\frac{1}{\overline{\mathcal{G}}}\left(\overline{\tilde{\mathcal{J}}^+\mathcal{G}}\right)^2+\right.\\
     \left.+(p')^2\left\langle\frac{1}{B^2}\right\rangle\frac{G}{\langle B^2\rangle}\right],
\end{multline}
\begin{equation}
        H=-\frac{p'}{G\Lambda(V')^2}\langle B^2\rangle\overline{\tilde{\mathcal{J}}^+\mathcal{G}}.
\end{equation}
The averaged magnetic energy can be straightforwardly written (assuming the flux surface average meaning of $\langle\dots\rangle$) in the form
\begin{equation}
    \langle B^2\rangle = \frac{4\pi^2G}{V'},
\end{equation}
and with the basic deinition for the volume $V=\int\mathrm{d}\psi\mathrm{d}\theta\mathrm{d}\phi \mathcal{J}$ with the appropriate limits, it follows that
\begin{equation}
    V'=4\pi^2\mathcal{J}_0.
\end{equation}
These constitute all the needed ingredients to write, putting all together, 
\begin{equation}
    D_R=\frac{p'\overline{\mathcal{G}}}{(\iota')^2}\left[\mathcal{J}_0p'\left\langle\frac{1}{B^2}\right\rangle-\mathcal{J}_0'\right]+\left(\frac{p'}{\iota'}\right)^2\overline{\mathcal{G}}\overline{\tilde{\mathcal{J}}^{+2}\mathcal{G}}.
\end{equation}
This is the form used in the main text Eq.~(\ref{eqn:Dr}). \par
We use this oportunity to point the differences between this form and that presented in [\onlinecite{hegna1989}]. Though the same Jacobian-based representation is used, the expressions differ in an additional $\mathcal{J}_0$ term. This term is a higher $\beta$ modification to the magnetic well piece of $D_R$, and is is missing in the results obtained in [\onlinecite{hegna1989}]. For the purpose of this paper, however, the arguments are not affected qualitatively by this additional piece.

\section*{Acknowledgements}
We thank Elizabeth Paul and Wrick Sengupta for stimulating discussions. This research is supported by grants from the Simons Foundation/SFARI (560651, AB) and DoE Contract No DE-AC02-09CH11466.

\section*{Data availability}
Data sharing is not applicable to this article as no new data were created or analyzed in this study.

\bibliography{currSingQS}

\end{document}